\documentclass{arxiv}
\usepackage{amssymb}
\usepackage{amsmath}
\usepackage{url}
\usepackage{subfigure}
\usepackage{rotating}
\usepackage{color}

\title{\bf Benford's law: a 'sleeping beauty' sleeping in the dirty pages of logarithmic tables}
\author{\small Tariq Ahmad MIR$^ {1,a}$ and Marcel AUSLOOS$^ {2,b}$\\
{\bf Accepted for publication in the Journal of the Association for Information Science and Technology}}

\address{\footnotesize
$^1$Nuclear Research Laboratory, Astrophysical Sciences Division, Bhabha Atomic Research Centre, Srinagar-190 006, Jammu and Kashmir, India. \\$^a$ $e$-$mail$ $address$: taarik.mir@gmail.com\\\
$^2$  GRAPES\footnote{Group of Researchers for Applications of Physics in Economy and Sociology} \\ rue de la Belle Jardini\`ere 483, B-4031, Angleur, Li\`ege, Belgium \\ 
$^b$$e$-$mail$ $address$:  marcel.ausloos@ulg.ac.be
}

\begin{document}

\catchline{}{}{}{}{}

\maketitle


\section{Abstract}
Benford's law 
is an empirical observation, first reported by Simon Newcomb in 1881 and then independently by Frank Benford in 1938: the first significant digits of numbers in large 
data are often distributed according to a logarithmically decreasing function. Being contrary to intuition, the law was 
forgotten 
as a mere curious observation. However, in the last two decades, 
relevant literature has grown exponentially, - an evolution typical of "Sleeping Beauties" (SBs) publications that go unnoticed (sleep) for a long time and then suddenly become center of attention (are awakened). Thus, in the present study, we show that Newcomb (1881) and Benford (1938) papers are clearly SBs. The former was in deep sleep for 110 years whereas the latter 
was in deep sleep for a comparatively lesser period of 31 years up to 1968, and in a state of less deep sleep for another 27 years up to 1995. Both SBs were awakened in the year 1995 by Hill (1995a). In so doing, we show that the waking prince  (Hill, 1995a) is more often quoted than the SB whom he kissed, - in this Benford's law case, wondering whether this is a general effect, -  to be usefully studied.
\section{Keywords}

Benford's law; sleeping beauty


\section{Introduction}	
 Amongst the enormous scientific literature that is being published annually, the reaction of the research community to the quality of a particular article, measured by the number of times it is being referred to by others, i.e. is cited, varies dramatically. Some articles may become instant hits, receive a large number of citations within the first few years of their publication, whereas most of the scientific articles are never cited at all (Seglen 1992), though the  proportion  of these has been decreasing over time (Larivi\'{e}re et al., 2008). Indeed, it has been shown that the number of citations of most articles, as a general rule, peaks with in the first few years of their publication (Gl\"{a}nzel \& Schoepflin, 1995; Gl\"{a}nzel et al., 2003; Costas et al., 2010) and thereafter, over the years, follow a decreasing trend till they receive none, and are then finally buried in the archives. However, there are some exceptions to this general evolution of citations. Articles that attract none or only few citations for a long time (sleep) since their publication and then suddenly their citation numbers burst (are awakened) are called Sleeping Beauties (SBs) (van Raan, 2004). Such articles achieve prominence following them being cited by some another article (the "prince"). In scientometrics, this phenomenon is also known under appellations like: resisted discoveries (Barber 1961), premature discoveries (Wyatt, 1961; Stent, 1972), delayed recognition (Cole, 1970; Garfield, 1990; Gl\"{a}nzel et al., 2003) and information awakening (Wang et al., 2012). On the other hand, articles which begin with a considerable number of citations in  the initial years after their  publication  can go to sleep for many years till they are again awakened by some prince; those are called "all-elements-sleeping-beauties" (Li \& Ye, 2012). Opposite to SBs are the flash-in-the-pans, articles which begin with many citations in the initial years of their publication but thereafter are forgotten completely (van Dalen \& Henkens, 2005). 
 
 The content of the paper, reputation of the author and his/her institute (Merton, 1968; Cole, 1970), impact factor of the journal - the percentage of uncited articles in high ranking journals is far less than those in the lower ranked journals (van Dalen \& Henkens, 2005), initial uncitedness - the longer an article takes to its first citation, the lesser the chance that it will ever be cited (Gl\"{a}nzel \& Garfield, 2004) and number of references - the more of them in the bibliography the more are the chances of article being cited (Stern, 1990) are  considered to  be factors which determine whether a given article will be a future SB. On the other hand, Burrell (2005) assumed the accumulation of citations to be purely random and showed that theoretically SBs can occur by chance without any need for the awakening prince or for any of special attributes discussed above.
 
Looking for a possible SB, in the sea of scientific literature, is a task no less than looking for a needle in the haystack. Nevertheless, scanning of entire citation data bases have thrown up contrasting results on the occurrence of SBs. van Raan (2004) concluded that SBs are indeed rare whereas Gl\"{a}nzel et al. (2003) and Ke et al. (2015) found the phenomenon not to be so exceptional. The contrasting results arise due to the different choice of threshold parameters like time of sleep and the depth of the sleep i.e. number of citations received during sleeping time. Thus longer the time of sleep and more the depth of sleep (lesser number of citations) the lesser the number of SBs (van Raan, 2004, 2015)) and vice-versa (Gl\"{a}nzel et al., 2003).  When unearthed SBs turn out to be application-oriented publications (van Raan, 2015) or having presented fundamental results (Gl\"{a}nzel et al., 2003), often ahead of their time, or are related to important discoveries, many later on winning Nobel Prize for the authors (Li and Shi, 2015). Three of the large scale studies, van Raan (2004, 2015) and Gl\"{a}nzel et al. (2003), analyzed the literature published after the year 1980 whereas Ke et al. (2015), more recent one, analyzed publications of American Physical Society from 1893 and that of Web of Science from 1900. Obviously any SBs older than the dates covered in these studies would be left out. Further, the citation practices vary among the disciplines (journals) and many potential SBs from a low citations discipline (journal) may slip through when analysis is done in conjunction with a discipline (journal) of higher citations. SBs from low citation journals (Kozak, 2013) and subjects of study (Ohba \& Nakao 2012) have been reported. See also: Abreu (2015)  or da Fonseca (2015). 

Individual papers of authors that went on to become SBs have also been pointed out. Van Calster (2012) reported a remarkable increase in the citations of Peirce (1984), a short note in Science. Marx (2014) showed that Shockley and Queisser (1961) is a notable example of a scientific SB. Gorry and Ragouet (2016) found that Dotter (1964), paper that lead to the birth of interventional radiology, is a SB. Below we introduce ``Benford's law (BL)'', an empirical observation that certainly was ahead of its time despite neither being any fundamental result nor being related to any prize winning discovery.   
  
 \section{Benford's law}
 
 That logarithmic tables begin with numbers having smaller initial digits is common knowledge. In 1881, a Canadian-American astronomer, Simon Newcomb, noticed that the first few pages of logarithmic table books are dirtier, more thumbed, than the latter ones. From this, he inferred that numbers with smaller initial digits are more often looked for and used than are the numbers with larger initial digits. Newcomb reported his observation as a two page note in the American Journal of Mathematics without providing any practical examples. It was forgotten for about six decades until its 1938 rediscovery by American physicist Frank Benford, apparently independent\footnote{Benford (1938) does not cite Newcomb (1881). In fact both  papers do not have any bibliography.} of the Newcomb (1881), through similar observation on the dirty pages of logarithmic tables.  Benford went much ahead in detail and tested the accuracy of his observation by analyzing a large collection of 20000 numbers he gathered from twenty diverse fields, thereby  establishing a law in form of the following mathematical equation. 
 \begin{equation}
P(d)=log_{10}(d+1)-log_{10}(d)=log_{10}(\frac{d+1}{d})=log_{10}(1+\frac{1}{d}), d= 1, 2, 3...,9
\end{equation}
where $P(d)$ is the probability of a number having the first non-zero digit d and $log_{10}$ is logarithmic to the base 10. Thus, in a given data set, the theoretical proportions for each of the digits from 1 to 9 to be first significant digit are those   shown in Table \ref{ta1}.
\begin{table}[h]
\tbl{The distribution of first significant digits as predicted by BL}
{\begin{tabular}{@{}ccccccccccc@{}} \toprule
Digit \hphantom{00} & 1 & 2 & 3 & 4 & 5 & 6 & 7 & 8 & 9 \\
Proportion \hphantom{00} & 0.301 & 0.176 & 0.125 & 0.097 & 0.079 & 0.067 & 0.058 & 0.051 & 0.046 \\ \botrule
\end{tabular} \label{ta1}}
\end{table} 
The phenomenon is now popularly only known as BL, a name questionable on account of its lack of acknowledgment of  Newcomb's contribution. Newcomb's law or Newcomb-Benford law would have been more appropriate. Nevertheless, the naming of the phenomenon is in accordance with Stigler's law of eponymy (Stigler, 1980).

The first significant digit of a number is its left-most nonzero digit. According to BL, in large data, the proportion of a digit as the first significant digit decreases logarithmically as the value of the digit increases. Thus, the  smallest digit, 1, will appear as the  first digit with the highest proportion (30\%), whereas the  largest digit, 9, will appear as first digit with the least proportion (4\%). This goes completely against the common thinking that the first digits of decimal numbers in large data ought to be distributed uniformly, with an equal proportion of about 11\%, irrespective of the digit value. 

Simply unconvincing for novices, the law is also challenging the wisdom of researchers. Many attempts have been made to prove the law, but even after more than a century of its discovery, there is no agreement on the mathematical foundations of the law (Gauvrit \& Delahaye,  2008, 2009;  Berger \&  Hill, 2011; Ausloos et al., 2015). The popular explanation, regularity and large spread  of the data, for the ubiquity of the phenomenon by Feller (1966) has been found to be inaccurate too (Berger \& Hill, 2010, 2011, 2015). BL has been shown to be the only scale-invariant digit law (Pinkham, 1961);  its base invariance has also been established (Hill, 1995b). Further, BL has been verified on data from an enormous range of processes (Mir, 2012, 2014; Mir et al., 2014). Yet its universal nature is shattered by some data sets whose digit distributions do not follow the law. There is no set of absolute rules one can use to predict, on inspection, which type of data will conform to the law and which will not. Some rules of thumb, arrived at from the experience of working with data over the years, are (a) the appearance of numbers in the data should be free from any human restrictions, (b) data must span several orders of magnitude, and (c) the number of records in the data must be large enough for the manifestation of individual digits. 

Notwithstanding the skepticism in earlier years of its discovery, BL has emerged as a subject matter of books, peer reviewed papers, popular magazine articles, patents, newspaper reports and blogs. According to Nigrini (2012) up to 1975 only 50 papers on BL were published and by 2000 there were about 150 papers. In other words only 100 papers were published in a period of 25 years, an average of four papers per year. Compare this with the present rate (Mir, 2016), of more than 50 papers being published every year.  The phenomenal growth of BL related research is interestingly driven by two of its contrasting peculiarities. Indeed; papers will keep pouring as long as there is no consensus on the mathematical foundations of the law. The simple mathematical form of the law, on the other hand, makes its application quite straightforward:  for testing its validity of BL, one essentially needs data plethora of which are readily available from Internet. 

Here below, we quantify the delayed recognition of BL through the use of bibliometric indicators. Specifically, the yearly number of citations received by Newcomb (1881) and Benford (1938) are analyzed. While doing so,  we only discuss the contributions: reviewing the vast literature is a formidable task, which from time to time have repositioned BL as a subject of practical usage requiring much academic rigor.  Nevertheless, it is immediately found that Newcomb (1881) was in deep sleep for 110 years, whereas the time  of deep sleep for Benford (1938) is 31 years only, -  the immediate appreciation of latter being attributed to its preceding an important physics paper by Bethe et al. (1938). 

We argue that the lack of acknowledgement of Newcomb (1881) in (all) the papers of prominent workers, - some of which 
are currently the most cited papers on BL, may have further contributed to its long deep sleep. The number of citations to both  papers started to increase significantly following the publications of Hill (1995a) and Nigrini (1996), two VERY important papers on BL: the former dealt with the mathematical explanation of the law, whereas   the latter provided the first practical application of law. Hill (1995a) emerges as the prince for both Newcomb (1881) and Benford (1938). 


\section{Data}
Work done on BL has been reviewed earlier, see Raimi (1976) and Nigrini (1994). Hurliman (2006) gathered 349 papers published on BL up to July 5, 2006. There is also an exhaustive compilation by Beebe (2016). Further on, as a quick reference for interesting  researchers,  the single online resource updated continuously and exclusively dedicated to the publications on mathematical aspects of the law and on its applications across multiple disciplines should be pointed out, and  is ``Benford Online Bibliography'' (BOB) (Hill et al., 2009). As on March 30, 2016 this database had indexed 976 articles. Articles can be arranged by the year of their publication and also according to the order of the alphabets of the respective first author names. Furthermore, it provides a complete reference and online information about the selected article, the articles citing a given articles and the articles a given article is citing within its own database.

\section{Benford's law applied to its own literature}

Though present study is not about testing the validity of BL, a curious reader might nevertheless ask how does the law fare against the numbers quantifying its own literature. Two bibliometric indicators, number of publications and their citations, immediately come to mind. Miller (2015) studied the yearly number of ``Benford relevant`` publications listed on BOB and found their first digit distribution to be in good agreement to BL. Next, Mir (2016) used Google Scholar (GS) to collect the data on the numbers of citations received by all the articles citing Newcomb (1881) and Benford (1938). He  found that the distribution of the first digits of citations data are in very good agreement to BL; this has remained consistently true over the years.

\section{Citation Curves of Newcomb (1881) and Benford (1938)}

To study the evolution of research on BL,  we arranged all the publications indexed on BOB in chronological order. The first publication on BL obviously is Newcomb (1881). We noted down the yearly number of articles published on BL. Many of the entries on BOB are found to be magazine articles, newspaper reports and blog items which usually do not have any bibliography. In line with our aim of studying the citation history of the BL, we visually inspected the reference list of each of the articles indexed on BOB to find out which ones really cite,  i.e. refer to, Newcomb (1881) and Benford (1938). For example 10 articles were published on BL in the year 1981, but none out of which cites Newcomb (1881), whereas Benford (1938) is cited by two texts only; another example: 15 publications are listed for year 1998, but  the former paper  is cited by only 5,  whereas the  latter is cited  by 9 of them.

\begin{table}[h]
\tbl{Rate of increase or decrease of the number of citations  and co-citations as a function of time for various cases as discussed in the text}
{\begin{tabular}{@{}ccccccccccc@{}} \toprule
case\hphantom{00} &  Newcomb & Benford & Hill & Hill-Newcomb & Hill-Benford   \\ \hline
increase\hphantom{00} & 0.158 & 0.142 & 0.245 & 0.224 & 0.215   \\
decrease \hphantom{00} & 0.077 & 0.246& 0.149 & 0.251 & 0.252  \\ \botrule
\end{tabular} \label{ta2}}
\end{table} 


The  (yearly) time dependence of the number of citations received by Newcomb (1881)  is plotted in Fig. 1. 
The increase in citation rate is found to be exponential with a regression coefficient such that $0.96\ge R^2\ge 0.91$; after reaching a peak,  the number of citations goes down relatively faster.  In the decay,   the regression coefficient is smaller due to the small number of data point.  see Table \ref{ta2}. 
This is similar to what is found in the literature either for  scientific citations (Della Briotta Parolo  et al., 2015) or sales (Lambiotte \& Ausloos, 2006).  In some sense, scientific reports through  journals is akin to  authors   "sale"  of  papers (first to editors, next to reviewers, and {\it in fine} to their community), - whence a possible   "attention  dependence  interpretation"  (van Dalen \& Henkens, 2005; Huberman et al., 2009) of the data citation history, in general and for  BL in particular. 
 
The yearly number of citations received by  Benford (1938) is also plotted in Fig. 1.  The characteristic features of a SB, i.e.,  no or very few citations for an initially long time, followed by a rather  sudden jump in citations, are clearly visible. Newcomb (1881) was cited for the first time in 1920 by Boring (1920), 38 years after its publication. Then it went on without any citations for another 55 years up to   the year 1975. It received two citations each in years 1976, 1978 and only one citation in 1979. Thus,  in the 100 year period, from 1881 to 1980, Newcomb (1881) had received only 6 citations. For the next ten years, from 1981 to 1990, it received in all 9 citations:  4 of these citations coming in the year 1984 alone. With an average of $\ll$ 1 citation per year Newcomb (1881) was in deep sleep for a massive 110 years. It continued to receive 1-digit citations per year from 1991 to 2000 leading to a total 37 citations, in this decade, i.e. 3.7 citations per year. Thus it is in this decade that Newcomb (1881) was awakened from its deep sleep. Next, we will show that this awakening is due to Hill (1995a). In the year 1995,  Newcomb (1881)  received a total of four citations out of which three citations alone came from the papers of Hill.

Benford (1938) received its first citation in 1944 after a silence gap of 6 years since its publication. Next it received citations (one  each) in 1948, 1961, 1966 and 1968. With a total of 5 citations up to 1968, Benford (1938) was in deep sleep for a period of 31 years. Citations jumped to 4 in 1969. The paper was in  a "less deep sleep" for another 7 years from 1969 to 1975, when it received an average of 1.85 (13/7) citations per year. The 1-digit citations to Benford (1938) continued up to year 1995 in which year it received 8 citations, -  out of which 3 came from Hill (1995a, 1995b, 1995c). In 1996, Benford (1938)  received 10 citations. Thus, Benford (1938) was also awakened by Hill (1995a). The sudden drop in the respective curves towards the end of the examined time interval,  can be attributed  to incomplete data  mining, for   2014, 2015 and 2016. Indeed, there is a certain delay between the publication of an article and its indexing on BOB. Further, on the date of accession, the online bibliography was shown to have been  "last update(d)" on May 1, 2015.

\begin{figure}
\hspace*{0pt}
\subfigure{\label{}\includegraphics[width=0.6\linewidth, height=1\linewidth,  angle=270,]{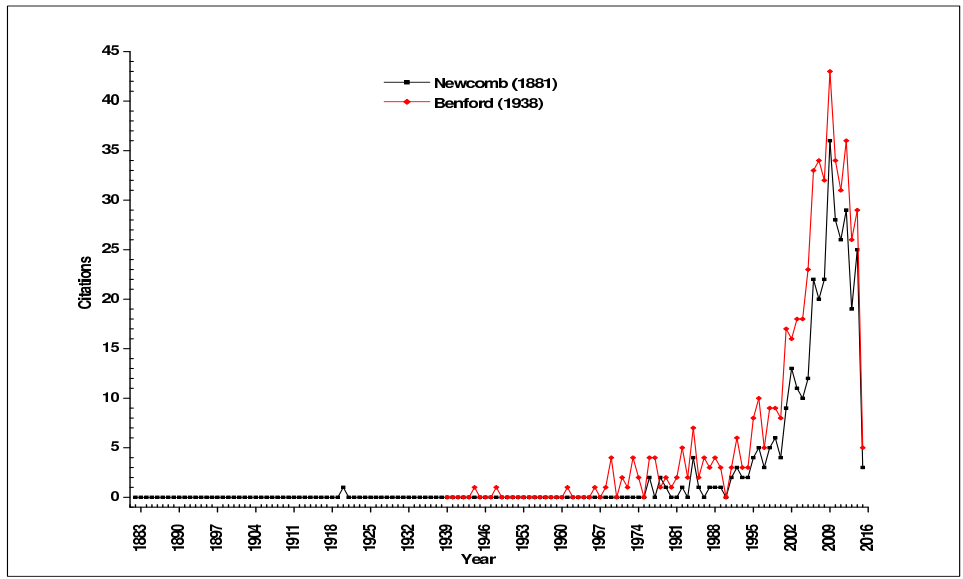}}
\vspace*{-30pt}
\caption{Yearly citations of Newcomb (1881) and Benford (1938).}
\end{figure}
The different evolution of citations of the two papers is further illustrated in Fig. 2 where the yearly number of papers citing (i) Newcomb (1881) but not Benford (1938) (N/B) (ii) Benford (1938) but not Newcomb (1881) (B/N) and (iii) of those citing both the papers (B \& N) are compared. As can be seen from the inset plot from 1938 to 1975 there are no citations to Newcomb (1881). Upto year 1995 the values of B/N are greater than both N/B and B \& N. In fact B/N exceeds N/B through out the citation history of the two papers indicating that whenever Newcomb (1881) is cited Benford (1938) is also cited but the reverse is not true. It is only after 1996 that B \& N exceeds B/N i.e. researchers increasingly begin to cite the two papers simultaneously thereby again reaffirming the fact that Hill (1995a) brought Newcomb (1881) back in focus.
\begin{figure}
\hspace*{0pt}
\subfigure{\label{}\includegraphics[width=0.6\linewidth, height=1\linewidth,  angle=270,]{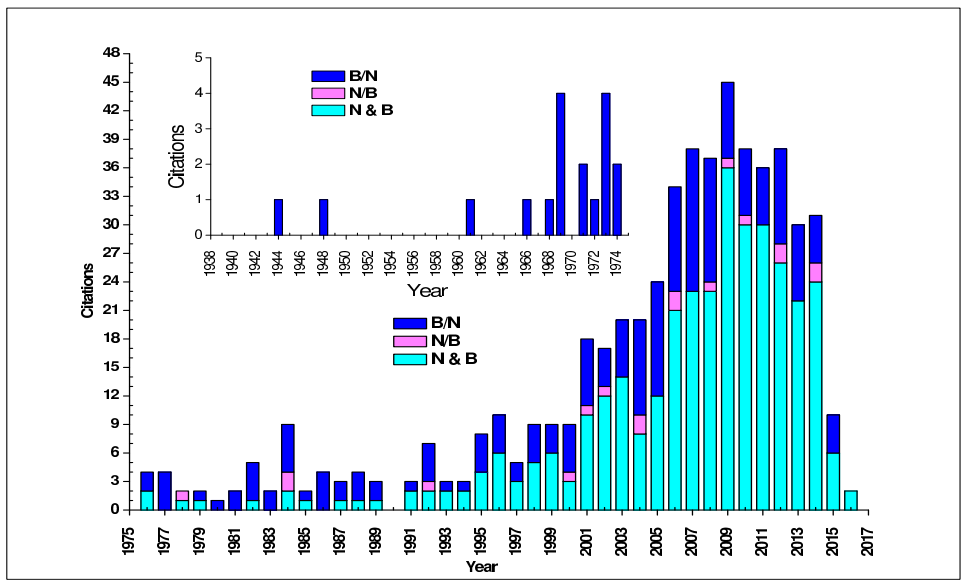}}
\vspace*{-30pt}
\caption{Comparison of yearly citations received by Benford (1938) only (B/N), Newcomb (1881) only (N/B) and both (B \& N). Inset plot is for years 1938-1974.}
\end{figure}
\subsection{Appreciation of Benford (1938)}
Goudsmith and Furry (1944) (32 citations) cited Benford (1938) for the first time attempting an explanation of the phenomenon with further expansion by Furry and Hurwitz (1945). When after 33 years Goudsmith (1977) revisited the problem 40 papers on BL had already appeared much contrary to his anticipation. Thus Goudsmith (1977) and Logan and Goudsmith (1978) attribute the quick pick up of Benford (1938) to its good fortune of preceding, Bethe et. al (1938), an important physics paper in the same volume and issue of the journal. None of these publications but Logan and Goudsmith (1978) cited Newcomb (1881). It is fascinating to note  that in terms of citations Benford (1938) has marched much ahead of Bethe et. al (1938). A GS search on May 3, 2016 showed that the former had 1114 citations whereas the latter has 289 citations only. 

The above insight behind the immediate pick up of Benford (1938) 
looks convincing when the present citations to Mitchell (1881) and Sylvester (1881)  articles preceding and succeeding Newcomb (1881) are taken into account. A GS search performed on May 3, 2016 showed only one citation  to the former and twenty citations to the latter paper. Thus the position of Newcomb (1881) in the journal was less fortunate. However, Newcomb (1881) did not receive the attention that would be expected,  strangely, even after the publication of papers that are currently the most cited papers on BL.

\subsection{Notable contributions}
\subsubsection{R. Pinkham}
In 1961,  Roger Pinkham worked out two most significant results on BL. The first is the scale-invariance of the law which means that the  law should apply irrespective of the units in which measurements of a given process are expressed. This implies that if the law holds for data on the area of rivers, i.e. the first sample  mentioned in Benford (1938), measured in square meters,  it will also hold if the measurements were to be expressed in square feet. In other words scale-invariance of BL implies its invariance under multiplication. If all the numbers in a sample of data obeying BL are multiplied by a non-zero constant, then the resulting sample will also obey the law. The second rule Pinkham (1961) worked out is that BL is the only scale-invariant distribution of first significant digits .

Pinkham (1961) did not cite Newcomb (1881) but cited only Benford (1938). With 190 citations Pinkham (1961) is the second most cited paper on BL. Newcomb (1881) did not receive a single citation for 15 years after Pinkham (1961) (Fig. 1 \& 2) till it was cited in another prominent paper (Raimi, 1976). During this period Benford (1938) received 16 citations out of which  it is co-cited  9 times with Pinkham (1961). From 1961 to 1995, Newcomb (1881) and Pinkham (1961) are co-cited only 9 times,  whereas the latter and Benford (1938) are co-cited 31 times. Pinkham (1961) is co-cited with Newcomb (1881) 132 times whereas with Benford (1938) it shares 173 citations. Thus,  a timely citation by Pinkham (1961) kept Benford (1938) in some limelight, whereas a lack of  consideration pushed Newcomb (1881) into oblivion.
\subsubsection{R. A. Raimi}
Raimi furthered the mathematical understanding of BL, using Banach and other scale invariances (Raimi, 1969a), and also its popularity (Raimi, 1969b). Both papers cite only Benford (1938) which received 4 citations in 1969 whereas none cited Newcomb (1881) which received no citation  at all in this year (Fig. 2). Raimi cited both Newcomb (1881) and Benford (1938) in his review of the relevant literature (Raimi, 1976) where he pointed out deficiencies in various available explanations. Raimi (1976) is currently the third most cited paper on BL with 174 citations. Up to 1995 Raimi (1976) is cited 46 times whereas Newcomb (1881) is cited 25 times. Overall Raimi (1976) is co-cited with Newcomb (1881) 128 times whereas it is co-cited with Benford (1938) 149 times.  Raimi (1985) again debated the genesis of the first digit phenomenon.
\subsubsection{T. P. Hill}

Following its citation by Pinkham (1961) and Raimi (1976) Benford (1938) kept receiving 1-digit citations up to 1995 (8 for this year). The first 2-digit citation to Benford (1938), 10, appeared in 1996;  afterwards  the number of citations   increased sharply. Around the same time, the citations of  Newcomb (1881), having received only 28 citations up to 1995, also started to increase. 
This is due to the publication of three papers by Hill (1995a, 1995b, 1995c),  all citing both Newcomb (1881) and Benford (1938) on the mathematical aspects of BL. Hill (1995a) proposed one of the most convincing explanations of BL,  demonstrating that if distributions are selected at random (in an ``unbiased'' way) and  if random samples are then taken from each of these distributions, the significant digits of the combined sample will converge to the logarithmic (Benford) distribution. With 305 citations Hill (1995a) is the most cited paper on BL. Hill (1995b) proved that base-invariance implies BL, i.e. another distinguishing and interesting  property of the law. 

Hill (1988) is the first publication by  Hill on BL.  However, it cites only Benford (1938) despite having two references, Raimi (1976) and Cohen (1976), in its bibliography which, besides Benford (1938) also cite Newcomb (1881). This paper is cited 39 times; it  reported that the initial digits of 6-digit numbers written by students in a class room experiment, though not in complete agreement to BL, were found more likely to be  the small  digits.
\subsubsection{M. J. Nigrini}
Hill's 1995 papers were  purely mathematical. BL really caught the fancy of the common people, auditors and academic researchers following wide spread media publicity of Nigrini's work who  advocated the first practical application of the law as a statistical tool for the detection of tax (fraud or) evasion. In his 1992 Ph. D thesis, (39 citations), Nigrini showed that BL can be used for the detection of manipulations of income tax returns filed by individuals to the Internal Revenue Services of USA. Nigrini cited both Newcomb (1881) and Benford (1938) in his Ph. D thesis. Afterwards Nigrini published a popular article Nigrini (1993), cited only 6 times, thereafter reviewing   developments on the understanding and applications of BL (Nigrini, 1994), cited 29 times. However, Nigrini's most acclaimed  paper, i.e. Nigrini (1996), also the third most cited paper on BL with 159 citations, again surprisingly  cited only Benford (1938) but not Newcomb (1881). In fact, Nigrini's second and third most cited papers (Nigrini \& Mittermaier, 1997; Nigrini, 1999) again do not cite Newcomb (1881). 

\subsection{Identifying the Prince}
In order to identify the prince for both Newcomb (1881) and Benford (1938) we adopt the procedure outlined in Braun et al. (2010): a candidate prince should be among the first citing articles which are highly cited and have a number of co-citations with the SB. The most cited papers on BL and the number of their co-citations with Newcomb (1881) and Benford (1938) are listed in Table \ref{ta3}. Pinkham (1961) and Nigrini (1996) are ruled out as possible prince candidates for Newcomb (1881) as it is cited by neither of the two. Newcomb (1881) is cited by both Raimi (1976) and Hill (1995a). However, the latter being the most cited paper on BL (305 citations), and co-cited with Newcomb (1881) 203 times, more frequently than is Raimi (1976), clearly emerges as the prince (Fig. 3). 

The citations to Benford (1938) started to increase after the publication of Hill (1995a) and Nigrini (1996), papers published around the same time. The citations of Hill (1995) and Benford (1938) are shown in Fig. 4. Hill (1995a) is co-cited with Benford (1938) 244 times whereas Nigrini is cited as 159 times and co-cited with Benford (1938) only 142 times. 
The citations to Hill (1995) peaked at 29 in 2008 whereas to Benford (1938) at 44 in 2009 and both were co-cited a maximum of 26 times in 2009. Hill (1995) emerges as the prince of the Benford (1938).

\begin{table}[h]
\tbl{Most cited papers on BL arranged in chronological order}
{\begin{tabular}{@{}ccccccccccc@{}} \toprule
Paper \hphantom{00} & Citations & Cites & co-cited & Cites  & co-cited  & \\
\hphantom{00} &  & Newcomb & with Newcomb & Benford & with Benford & \\ \botrule
Pinkham (1961) \hphantom{00} & 190 & No & 132 & Yes & 173 & \\
Raimi (1976) \hphantom{00} & 174 & Yes & 128 & Yes & 149 & \\
Hill (1995) \hphantom{00} & 305 & Yes & 203 & Yes & 244 & \\
Nigrini (1996) \hphantom{00} & 159 & No & 115 & Yes & 142 & \\
 \botrule
 \hphantom{00} & Citations to Newcomb (1881) & 366 & Citations to Benford (1938) & 550 & \\
 \botrule
\end{tabular} \label{ta3}}
\end{table}

\begin{figure}
\hspace*{0pt}
\subfigure{\label{}\includegraphics[width=0.6\linewidth, height=.98\linewidth,  angle=270,]{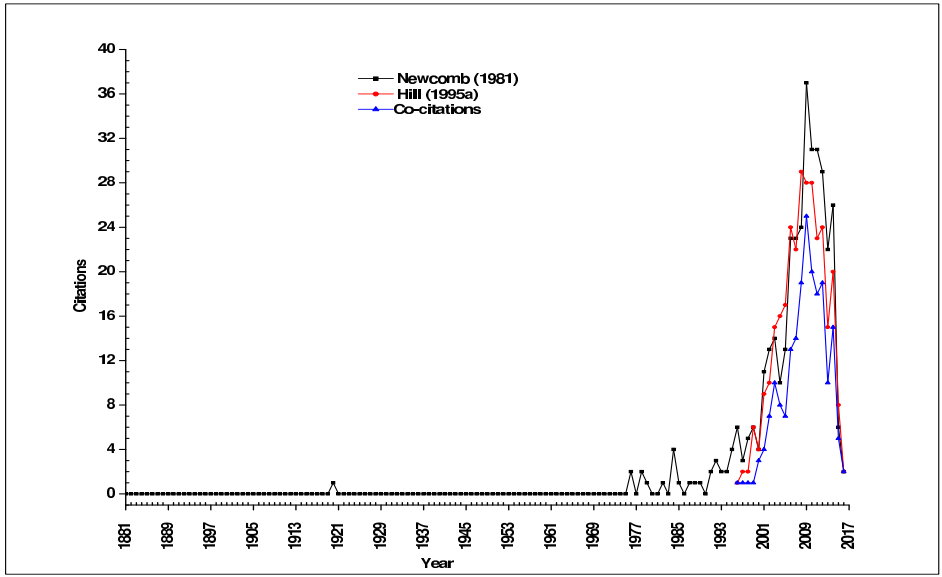}}
\vspace*{-30pt}
\caption{Yearly citations of Newcomb (1881) and Hill (1995a).}
\end{figure}

\begin{figure}
\hspace*{0pt}
\subfigure{\label{}\includegraphics[width=0.5\linewidth, height=.85\linewidth,  angle=270,]{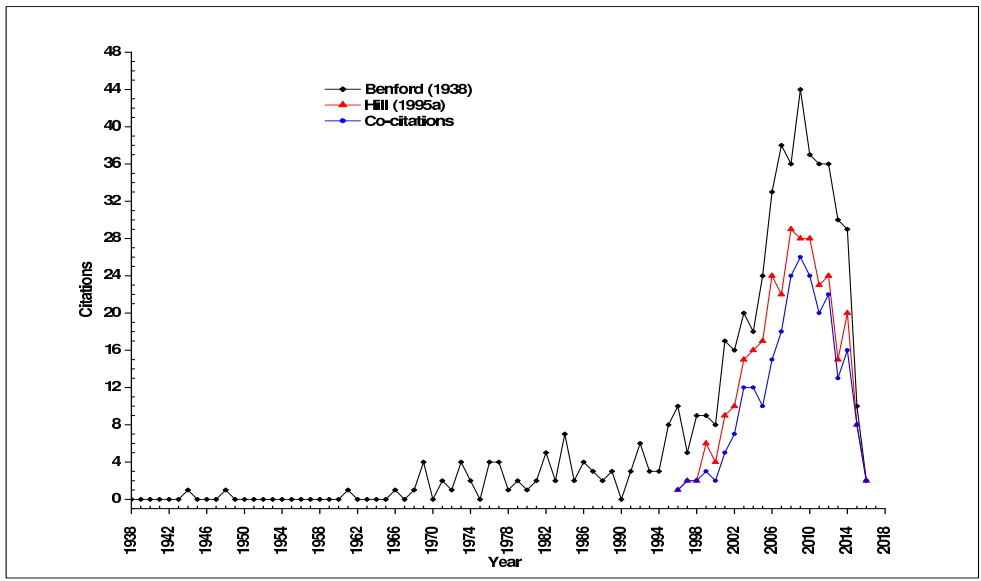}}
\vspace*{5pt}
\caption{Yearly citations of Benford (1938) and Hill (1995a).}
\end{figure}

\section{Conclusion}
It had therefore been demonstrated that Newcomb (1881) and Benford (1938) are  SBs. The former had a long deep sleep of 110 years whereas the latter, fortuitously preceding an important physics paper by Bethe et al. (1938), was in deep sleep for a comparatively lesser period of 31 years. The long dormancy of Newcomb (1881) and far lesser number of its citations, than Benford (1938), may have been fueled by an incomprehensible lack of its acknowledgement by prominent workers in all of their papers, - some of which are currently the most cited papers on BL.  Both SBs were woken up in 1995 by the prince, Hill (1995a), now the most cited paper on BL,  with maximum number of co-citations, amongst the most cited papers, with Newcomb (1881) and Benford (1938). 
The mid 1990s resurgence of BL is coincident with the growth of online databases. Indeed the validation of the law for data from a variety of sources forms a considerable size of Benford literature thereby making it a highly interdisciplinary field of research.

The variable influence of articles preceding and succeeding Newcomb (1881) and Benford (1938) on the accumulation of their future citations, markedly different, is in accordance with their contrasting importance. Hence a new direction of research in unearthing the potential SBs is posited i.e. whether the position of a SB in the journal has had any role in its awakening?\footnote{ For a recent discussion on how a paper's position in the issue of the journal affects its citations see Kosteas (2015).} This is particularly pertinent to the older articles which researchers, in the absence of digital libraries, had to read in the paper bound volumes of journals. Presently scientific literature is accessed online and one is seldom aware of the articles between which article of ones interest in the journal is sandwiched.  

The analysis of data which Benford  gathered over several years must have taken him considerable time, in the absence of computers; moreover,  he did all the calculations by hand. Today,  the analysis of the same information, invariably accessed online, can be done in a very short time. Furthermore, people hardly go to traditional libraries to look at the logarithmic table books,  and thus one would not find any dirty pages. Benford was rightly placed in time when he stumbled upon the dirty pages of logarithmic tables. Had it been the era of computers there would have been no BL. 

After their initial report, both Newcomb and Benford did not pursue the first digit phenomenon any further. They had other works on hand:  the former had the honor of presiding over various American scientific associations,  whereas  the latter besides being a recipient of 20 patents on optical devices has  published 109 papers in physics and mathematics.  Perhaps both scientists  anticipated that their papers, like the majority of others, would soon fade into obscurity. However, going by the current spurt in activity surrounding BL,  it is safe to claim that there is no other scientific discovery than BL,  with such a humble beginning and yet \textit{raison  d'\^etre}, for the popularity of its proponents.
 
\section*{Acknowledgments}
 MA  would like to thank R. Cerqueti, C. Herteliu, and B.V. Ileanu, for discussions on BL.

\end{document}